\documentclass[10pt,conference]{IEEEtran}
\IEEEoverridecommandlockouts
\usepackage{algorithmic}
\usepackage{algorithm}
\usepackage{bm}
\usepackage{cite}
\usepackage{amsmath,amssymb,amsfonts}
\usepackage{algorithmic}
\usepackage{graphicx}
\usepackage{textcomp}
\usepackage{xcolor}
\usepackage{makecell}
\usepackage{subfigure}
\usepackage{tcolorbox}
\usepackage{balance}
\usepackage{amssymb}
\usepackage{pifont}

\usepackage{multirow}
\usepackage{diagbox}
\usepackage{graphicx}
\usepackage{booktabs}

\usepackage{xcolor}
\usepackage{hyperref}

\definecolor{citegreen}{RGB}{0,255,0}

\usepackage{etoolbox}
\makeatletter
\patchcmd{\@cite}
  {\hbox{[#1]}}
  {\fboxsep=1pt\colorbox{citegreen}{\hbox{[#1]}}}
  {}
  {}
\makeatother
\def\BibTeX{{\rm B\kern-.05em{\sc i\kern-.025em b}\kern-.08em
    T\kern-.1667em\lower.7ex\hbox{E}\kern-.125emX}}
\begin{document}

\title{Smart-LLaMA: Two-Stage Post-Training of Large Language Models for Smart Contract Vulnerability Detection and Explanation}


\title{Smart-LLaMA: Two-Stage Post-Training of Large Language Models for Smart Contract Vulnerability Detection and Explanation\\
\thanks{$^*$These authors contributed equally to this work.}
\thanks{$^{\S}$Corresponding author: Li Yang}
}

\author{
  \IEEEauthorblockN{
      Lei Yu$^*$\IEEEauthorrefmark{2}\IEEEauthorrefmark{3}, 
      Shiqi Cheng$^*$\IEEEauthorrefmark{2},
      Hang Yuan\IEEEauthorrefmark{2}\IEEEauthorrefmark{3}, 
      Peng Wang\IEEEauthorrefmark{2}\IEEEauthorrefmark{3}, 
      Zhirong Huang\IEEEauthorrefmark{2}\IEEEauthorrefmark{3},
      Jingyuan Zhang\IEEEauthorrefmark{2}\IEEEauthorrefmark{3},\\
      Chenjie Shen\IEEEauthorrefmark{2}\IEEEauthorrefmark{3},
      Fengjun Zhang\IEEEauthorrefmark{2}, 
      Li Yang$^{\S}$\IEEEauthorrefmark{2},
      Jiajia Ma\IEEEauthorrefmark{2}
    }
  \IEEEauthorblockA{
    \IEEEauthorrefmark{2}Institute of Software, Chinese Academy of Sciences, Beijing, China \\
    \IEEEauthorrefmark{3}University of Chinese Academy of Sciences, Beijing, China
  }
}

\maketitle

\begin{abstract}

With the rapid development of blockchain technology, smart contract security has become a critical challenge. However, existing smart contract vulnerability detection methods face three main issues: (1) Insufficient quality and comprehensiveness of datasets, due to the lack of detailed explanations and precise vulnerability locations in current datasets. (2) Limited adaptability of large language models (LLMs) to the smart contract domain, because most LLMs are typically pre-trained on vast amounts of general text data but very little smart contract-specific data. (3) Lack of high-quality explanations for detected vulnerabilities, as most existing methods focus solely on detection without providing clear explanations for their results. These limitations significantly hinder detection performance and make it harder for developers to understand and fix vulnerabilities quickly, potentially leading to severe financial losses. To address these problems, we propose Smart-LLaMA, an advanced detection method based on the LLaMA language model. First, we construct a comprehensive dataset covering four vulnerability types with labels, detailed explanations, and precise vulnerability locations. Second, we introduce Smart Contract-Specific Continual Pre-Training, using raw smart contract data to enable the LLM to learn smart contract syntax and semantics, thereby enhancing their adaptability to the smart contract domain. Furthermore, we propose Explanation-Guided Fine-Tuning, a novel approach that fine-tunes the LLM using paired vulnerable code and explanations, enabling it to both detect vulnerabilities and provide reasoned explanations for its results. To evaluate the quality of generated explanations, we employ both LLM evaluation and human evaluation, focusing on three key aspects: Correctness, Completeness, and Conciseness. Experimental results show that Smart-LLaMA outperforms state-of-the-art baselines, with average improvements of 6.49\% in F1 score and 3.78\% in accuracy, while providing reliable explanations. We have made all models, datasets, and code available.







\end{abstract}


\section{Introduction}

The advent of blockchain technology has seen rapid adoption across various sectors, driven by its decentralized architecture \cite{swan2015blockchain}. This innovative technology enables the creation of secure, distributed digital ledgers for recording transactions \cite{hewa2021survey}. Utilizing advanced cryptographic methods, blockchain ensures the integrity and verification of each transaction, establishing itself as a highly reliable technological framework \cite{wood2014ethereum, yu2023money}. Within this ecosystem, smart contracts function as self-executing programs on the blockchain, automating the management of digital assets such as cryptocurrencies. These contracts activate when specific conditions are met and, once deployed, become permanent fixtures on the blockchain \cite{zou2019smart}. However, the immutable nature and inherent complexity of smart contracts present significant security challenges \cite{zou2019smart}. The well-documented DAO incident \cite{dhillon2017dao,mehar2019understanding} serves as a cautionary tale, illustrating the potential severity of such vulnerabilities. This security breach resulted in the unauthorized diversion of Ethereum valued at \$60 million, causing widespread disruption within the blockchain community \cite{alharby2017blockchain, hegedHus2018towards}. This event underscores the critical importance of enhancing smart contract security to prevent similar devastating outcomes in the future.

Researchers have developed various techniques to identify vulnerabilities in smart contracts, each addressing different aspects of the challenge but also facing limitations. Symbolic execution tools like Oyente \cite{luu2016making}, Mythril \cite{mueller2017mythril}, Osiris \cite{torres2018osiris}, and Manticore \cite{mossberg2019manticore}, as well as static analysis tools such as Slither \cite{feist2019slither} and SmartCheck \cite{tikhomirov2018smartcheck}, rely on predefined patterns to detect vulnerabilities. However, these methods often struggle with complex scenarios and lack generalizability. We conducted a detailed survey of existing smart contract vulnerability datasets as shown in \uppercase\expandafter{\romannumeral1}, evaluating multiple datasets including A \cite{wu2021peculiar}, B \cite{zhuang2020smart}, C \cite{yu2023pscvfinder}, and D \cite{qian2023cross}, and found significant limitations. These datasets typically provide only basic vulnerability labels, lacking detailed explanations and precise location information. They cover a limited range of vulnerability types, usually only 1 to 3, failing to represent the diverse potential security risks in smart contracts. This simplified labeling approach severely constrains models' ability to comprehensively understand and detect complex vulnerability patterns. These limitations directly affect the learning effectiveness of detection models, potentially leading to questionable accuracy and reliability in detection results.

Some more advanced methods have attempted to address these limitations. Clear \cite{chen2024improving} employs a Contrastive Learning (CL) model to capture complex inter-contract relationships, while Zhuang et al. \cite{zhuang2020smart} and Luo et al. \cite{luo2024scvhunter} introduce graph neural network-based approaches to represent smart contracts. However, the complexity of these graph structures makes them difficult to reproduce and less effective in representing programs accurately. Peculiar \cite{wu2021peculiar} and PSCVFinder \cite{yu2023pscvfinder} take a different approach by fine-tuning pre-trained models for vulnerability detection. While innovative, these methods still struggle to provide clear explanations for their detections, which is crucial for practical usage. Given these limitations, researchers have begun to explore the potential of using general-purpose Large Language Models (LLMs) to address smart contract vulnerability detection issues. General-purpose LLMs show promise in adapting to new patterns \cite{pal2024domain,naveed2023comprehensive}. However, they often struggle with smart contract-specific concepts and security implications. As illustrated in Figure 1, when presented with a smart contract, a general-purpose LLM like LLaMA-3.1-8B-Instruct incorrectly identifies a non-existent reentrancy vulnerability. It misinterprets the implications of external calls in the 'gotake()' function, failing to recognize that reentrancy vulnerabilities typically arise when contract state or balance changes occur after external calls, which is not the case in this contract.

To address these challenges, we propose our Smart-LLaMA, built upon the LLaMA-3.1-8B model. To overcome the limitations of existing datasets, we construct a comprehensive smart contract vulnerability dataset with detailed explanations and precise location information, covering four vulnerability types. This dataset is constructed through a three-step process: automated generation, LLM-based evaluation, and human expert verification and refinement. Specifically, we utilize the largest parameter versions of state-of-the-art LLMs (Qwen2 and Mistral-Large) to generate detection results, explanations, and specific vulnerability locations through carefully designed prompts. Llama-3.1-70B-Instruct serves as a judge model, evaluating these explanations on correctness, completeness, and conciseness. It scores each aspect from 1 to 10 to select the highest-quality explanations. Finally, human experts review the selected high-scoring explanations, verify their accuracy and make necessary improvements. This approach addresses the issue of insufficient dataset quality and comprehensiveness in existing resources. Furthermore, we introduce Smart Contract-Specific Continual Pre-Training to enhance the model's understanding of smart contract-specific syntax structures and vocabulary, thereby improving the adaptability of LLaMA-3.1-8B to the smart contract domain. This process involves exposing the model to a large corpus of original smart contract code, allowing it to learn the nuances and intricacies of smart contract development. Additionally, we propose Explanation-Guided Fine-Tuning, a novel approach utilizing our constructed smart contract vulnerability explanations to fine-tune the large language model. This process enables the model to comprehend the entire vulnerability detection process. By training on datasets pairing vulnerable code with detailed explanations, Smart-LLaMA learns to both identify vulnerabilities and articulate the reasoning behind its detections. To evaluate the quality of explanations generated by our Smart-LLaMA, we utilize both LLM evaluation and human evaluation. Our evaluation is based on three key dimensions: Correctness, Completeness, and Conciseness, each scored on a 4-point Likert scale \cite{joshi2015likert}. For LLM evaluation, we utilize Llama-3.1-70B-Instruct, carefully designing prompts to guide the model in assessing explanations based on these criteria. For human evaluation, we invite four experienced smart contract security experts. Each expert dedicate 8 hours to the assessment process, resulting in a total of 32 hours of in-depth analysis. The experts use the same 4-point Likert scale \cite{joshi2015likert}. To ensure consistency, we arrange for 20\% overlapping evaluation samples. We then tabulate the number of explanations receiving each score (1-4) for each dimension, providing a clear distribution of the quality assessments for both the baseline (LLaMA-3.1-8B-Instruct) and our Smart-LLaMA approach.

We evaluated our Smart-LLaMA framework on a challenging dataset \cite{qian2023cross} encompassing four major vulnerability types: reentrancy, timestamp dependency, integer overflow/underflow, and delegatecall. The results demonstrated that Smart-LLaMA significantly outperformed state-of-the-art methods across all vulnerability types. Notably, Smart-LLaMA achieved F1 scores 7.35\%, 1.24\%, 7.82\%, and 9.55\% higher for reentrancy, timestamp dependency, integer overflow/underflow, and delegatecall vulnerabilities compared to the previous best performers. In terms of accuracy, Smart-LLaMA surpassed the previous SOTA methods by 4.14\%, 0.62\%, 4.83\%, and 5.53\% for these four vulnerability types. 


In addition to detection performance, we evaluated the quality of vulnerability explanations generated by Smart-LLaMA. Both LLM evaluation and human evaluation demonstrated that Smart-LLaMA produced more accurate, comprehensive, and concise explanations compared to LLaMA-3.1-8B-Instruct. For instance, in the human evaluation, Smart-LLaMA achieved the highest score (4 out of 4) for correctness, completeness, and conciseness in 69.5\%, 57.1\%, and 65.6\% of cases, respectively, significantly outperforming the baseline method.

The main contributions of this paper are as follows:
\begin{itemize}





\item We propose Smart-LLaMA, a novel method combining smart contract-specific pre-training and explanation-guided fine-tuning for smart contract vulnerability detection, achieving state-of-the-art performance on four main vulnerability types.
  
\item We construct a high-quality smart contract vulnerability dataset that not only provides label, but also includes detailed vulnerability explanations, overcoming the limitations of existing datasets.
  


\item To the best of our knowledge, we are the first to explore explanation quality in smart contract vulnerability detection. We validate Smart-LLaMA's effectiveness in generating high-quality explanations through both LLM evaluation and human evaluation.

\end{itemize}

We have made all source code and datasets utilized in this research available to the public at https://zenodo.org/records/13860344

\section{Background and Motivation}

\subsection{Problem Statement}

We propose an automated approach to detect vulnerabilities in smart contracts. Our method assigns a label $\hat{y}$ to each independent smart contract, where $\hat{y}=1$ indicates the presence of a vulnerability and $\hat{y}=0$ denotes security. Notably, our approach not only automatically identifies vulnerabilities but also provides detailed explanations for each detected vulnerability, including its type, location, and potential impact. We focus on four key vulnerability types: Reentrancy, Timestamp Dependency, Integer Overflow/Underflow, and Delegatecall.

\textbf{Reentrancy vulnerability} occurs when a contract calls an external contract or sends Ether before completing all necessary internal state changes. An attacker can exploit this vulnerability by repeatedly calling the vulnerable function before the original call is completed, potentially leading to unexpected behavior such as multiple withdrawals of funds.

\textbf{Timestamp dependence vulnerability} occurs when smart contracts rely on block timestamps for critical operations. Miners can manipulate these timestamps, potentially compromising contract integrity and leading to financial losses. This vulnerability often affects contracts using timestamps for random number generation or key decision-making processes.

\textbf{Integer Overflow/Underflow} occurs when the result of an arithmetic operation exceeds the storage range of the variable. In an overflow, the value "wraps around" to the minimum value for that type, while in an underflow, it "wraps around" to the maximum value. This can lead to unexpected contract behavior such as incorrect balances or out-of-control loops.

\textbf{Delegatecall} is a low-level function call that allows a contract to dynamically load code from another contract. While this provides powerful upgradeability, it can lead to severe security vulnerabilities if used improperly. The main risk is that the called contract executes in the context of the calling contract and can thus modify the calling contract's storage.

\textbf{We primarily focus on these four vulnerabilities for the following reasons:} (i) Empirical evidence shows that approximately 70\% of financial losses in Ethereum smart contract attacks are attributed to these vulnerabilities \cite{chen2020survey}. (ii) Existing works \cite{chen2020survey,gao2019easyflow,praitheeshan2019security} demonstrate that these vulnerabilities occur with higher frequency in Ethereum smart contracts compared to others. (iii) These vulnerabilities have significant impacts on contract security and functionality, potentially leading to severe economic losses. (iv) Despite being well-known, these vulnerabilities are still often overlooked or misunderstood due to the complexity and immutability of smart contracts.




\subsection{Motivations}

In this section, we analyze three key motivations behind our research on improving smart contract vulnerability detection. We use real-world examples and existing datasets to illustrate the limitations of current approaches. These motivations highlight the importance of high-quality datasets, domain-specific model adaptation, and explainable detection results in smart contract vulnerability detection.

\begin{table}[h]
\caption{Comparison of Smart Contract Vulnerability Datasets}
\centering
\begin{tabular}{|l|c|c|c|c|}
\hline
\textbf{Dataset} & \textbf{Label} & \textbf{Explanation} & \textbf{Location} & \textbf{Types} \\
\hline
Dataset A \cite{wu2021peculiar} & $\checkmark$ & \ding{55} & \ding{55} & 1 \\
\hline
Dataset B \cite{zhuang2020smart} & $\checkmark$ & \ding{55} & \ding{55} & 2\\
\hline
Dataset C \cite{yu2023pscvfinder}& $\checkmark$ & \ding{55} & \ding{55} & 3\\
\hline
Dataset D \cite{qian2023cross} & $\checkmark$ & \ding{55} & \ding{55} & 4\\
\hline
Our Dataset & $\checkmark$ & $\checkmark$ & $\checkmark$ & 4\\
\hline
\end{tabular}
\label{tab:dataset_comparison}
\end{table}

\textbf{Motivation 1: Insufficient Quality and Comprehensiveness of Datasets.} We conducted a systematic evaluation of existing smart contract vulnerability datasets. As shown in Table I, although Datasets A \cite{wu2021peculiar}, B \cite{zhuang2020smart}, C \cite{yu2023pscvfinder}, and D \cite{qian2023cross} all provide vulnerability labels, they lack detailed vulnerability explanations and specific location information. More importantly, most datasets cover a limited number of vulnerability types (ranging from 1 to 3). This simplistic labeling approach and limited vulnerability type coverage severely constrain the model's ability to comprehensively understand and detect vulnerabilities. The inadequacy of dataset quality directly affects the learning effectiveness of models, leading to questionable accuracy and reliability of detection results. Moreover, the absence of detailed explanations and location information hampers the model's ability to learn precise detection patterns, potentially leading to overfitting on simplistic features and reducing the overall effectiveness and generalizability of the vulnerability detection process.

\begin{figure}[htp]
\centerline{\includegraphics[width=0.5\textwidth]{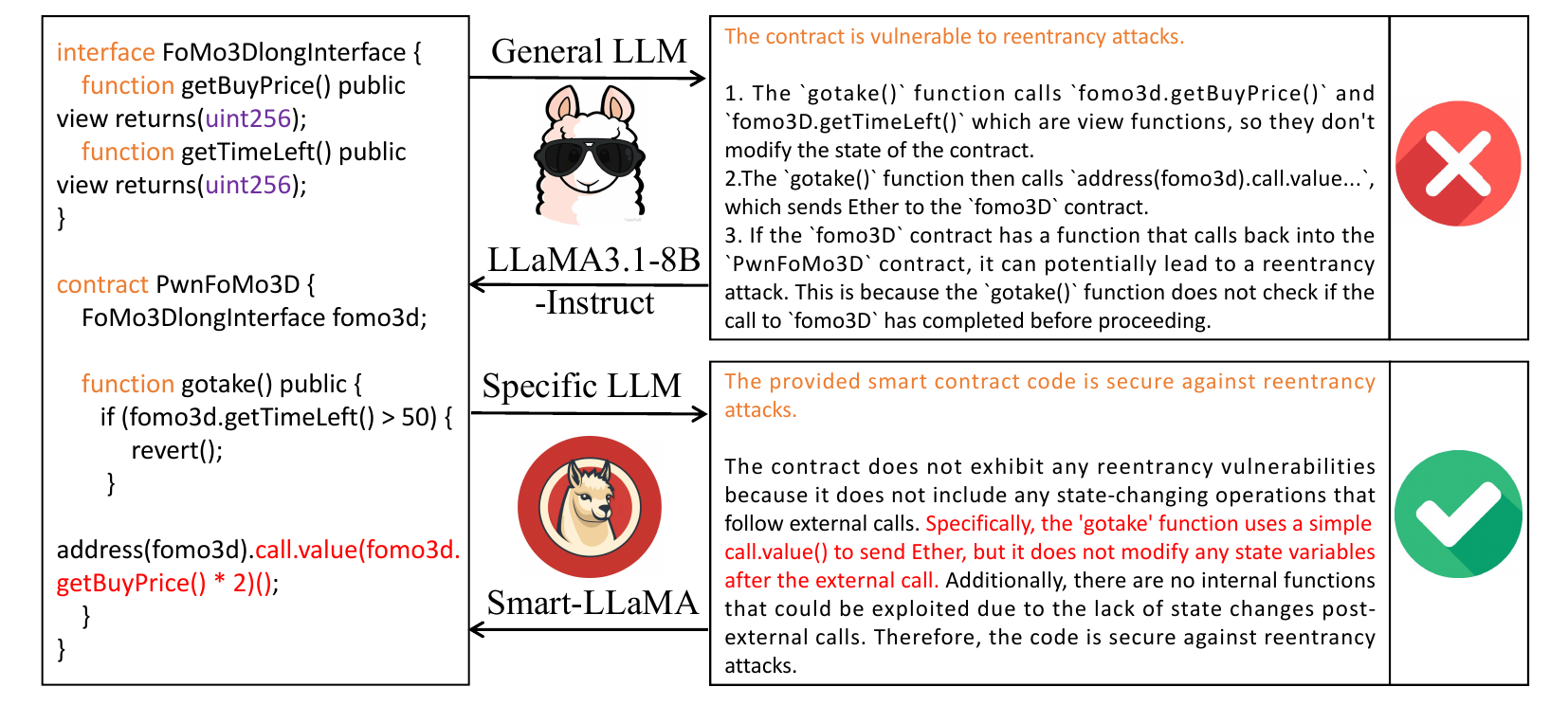}}
\caption{A Motivation Example of Smart Contract Vulnerability Detection with and without domain-specific post-training.}
\label{fig}
\end{figure}


\textbf{Motivation 2: Insufficient Adaptability of General Large Language Models in the Smart Contract Domain.}  General LLMs (such as LLaMA-3.1-8B-Instruct) demonstrate significant limitations when analyzing smart contracts. General LLMs, while proficient in general tasks, often misinterpret crucial smart contract concepts and security implications as shown in Figure 1. In this example, the general LLM incorrectly identifies a non-existent reentrancy vulnerability, misunderstanding the implications of external calls in the 'gotake()' function. It fails to recognize that reentrancy vulnerabilities typically arise when contract state or balance changes occur after external calls, which is not the case in this contract. This lack of domain-specific knowledge leads to misinterpretation of smart contract functionality and false identification of security vulnerabilities. In contrast, specialized LLMs like our Smart-LLaMA accurately assess the contract's security, correctly identifying that the code is secure against reentrancy attacks due to the absence of state changes after external calls. This stark difference highlights the critical need for domain-specific training in smart contract analysis to avoid potentially catastrophic misjudgments in real-world applications.

\textbf{Motivation 3: Lack of Explanation for Detecting Results.} Most existing methods focus solely on vulnerability detection results, neglecting to provide explanations for these results. In practical applications, merely knowing that a vulnerability exists is far from sufficient. Developers need to understand the nature of the vulnerability, its causes, potential impacts, and possible attack vectors to implement appropriate fixes. The lack of explainability not only reduces the credibility of detection results but also increases the complexity and cost of the fix process. As shown in the dialogue in Figure 2, developers need to understand not only the existence of vulnerabilities but also "Why?" and "Where?" to truly grasp the essence of the vulnerability. Furthermore, the absence of effective evaluation methods to assess the quality of generated vulnerability explanations makes it difficult to ensure that the generated explanations are both accurate and practical.

\begin{figure}[htp]
\centerline{\includegraphics[width=0.5\textwidth]{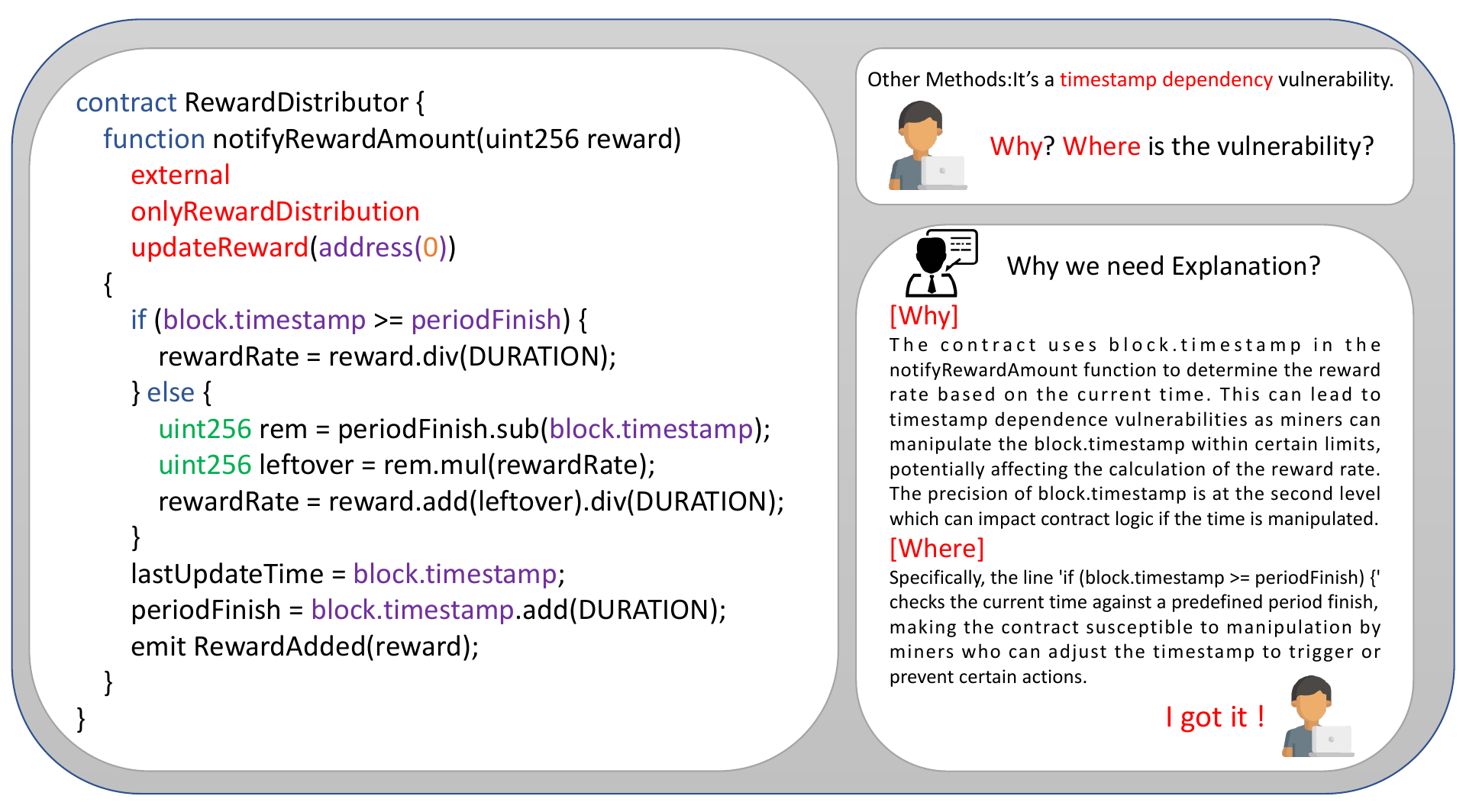}}
\caption{A Motivation Example of Smart Contract Vulnerability Detection with and without explanation.}
\label{fig}
\end{figure}

To address these challenges, we propose our Smart-LLaMA, built upon the LLaMA-3.1-8B model. First, to overcome the limitations of existing datasets, we construct a comprehensive smart contract vulnerability dataset with detailed explanations and precise location information, covering four vulnerability types. This addresses the issue of insufficient dataset quality and comprehensiveness (Motivation 1). Second, we introduce Smart Contract-Specific Continual Pre-Training to enhance the model's understanding of smart contract-specific syntax structures and vocabulary, thereby improving the adaptability of LLaMA-3.1-8B to the smart contract domain (Motivation 2). Finally, we propose Explanation-Guided Fine-Tuning to train the model to simultaneously perform vulnerability detection and generate high-quality explanations, addressing the lack of explanation for detecting results (Motivation 3). We also employ LLM evaluation and human evaluation to assess the quality of explanations generated by our Smart-LLaMA. 


\section{Approach}

\begin{figure*}[htp]
\centerline{\includegraphics[width=0.9\textwidth]{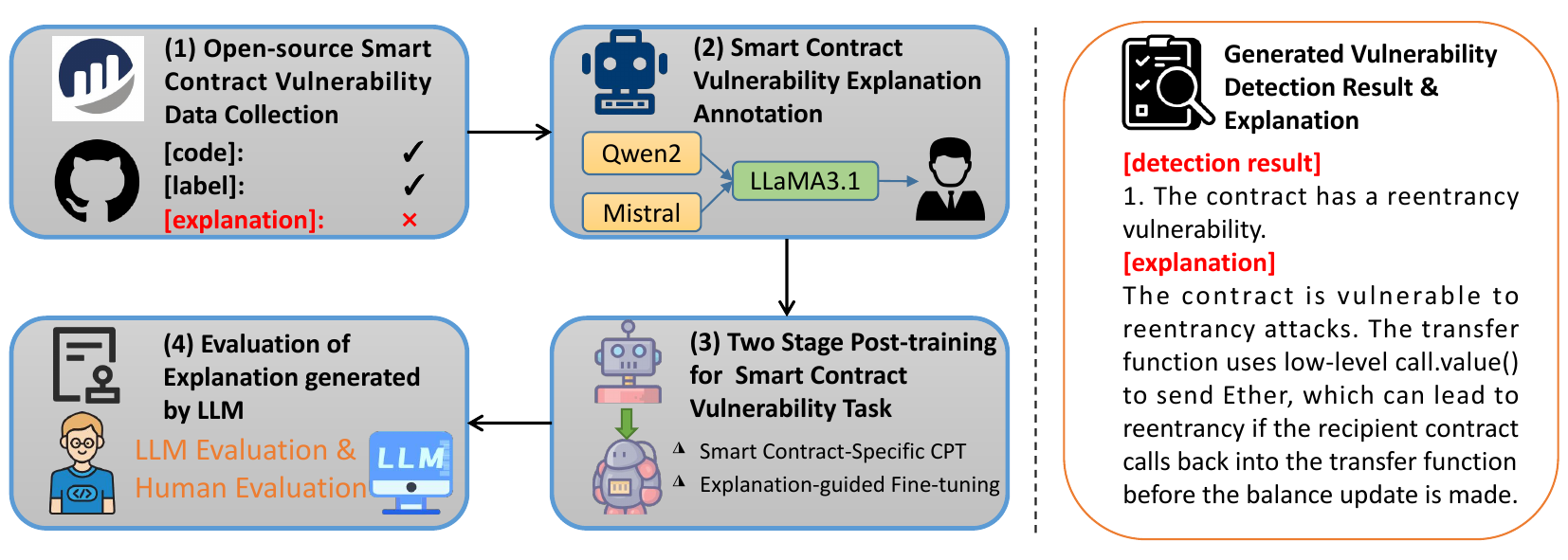}}
\caption{The Overview of our Smart-LLaMA.}
\label{fig}
\end{figure*}

The overall workflow of Smart-LLaMA is shown in Figure 3. The Smart-LLaMA framework consists of four key modules: Open-source Smart Contract Vulnerability Data Collection, Smart Contract Vulnerability Explanation Annotation, Two Stage Post-training for Smart Contract Vulnerability Detection Task, and Evaluation of Explanation generated by LLM.

\subsection{Open-source Smart Contract Vulnerability Data Collection} 

\subsubsection{Smart Contract-Specific Continual Pre-training}

Our Smart Contract-Specific Continual Pre-training dataset is derived from the work presented in \cite{storhaug2023efficient}. This dataset underwent rigorous filtering and quality validation processes. Specifically, Google BigQuery was utilized to select all smart contract addresses on the Ethereum blockchain with at least one transaction. Subsequently, the researchers queried Etherscan, the largest verified smart contract provider, to obtain the source code of these addresses. 

To ensure uniqueness, the researchers employed the computationally efficient token-based similarity algorithm, Jaccard Index \cite{allamanis2019adverse}. The downloaded smart contracts contained numerous duplicate library codes. To mitigate this, each contract file was decomposed into its original representative files. Library code, along with other imported contract files, was separated into individual contract records. The resulting smart contracts were then grouped by filename and filtered for uniqueness using a Jaccard Index \cite{allamanis2019adverse} similarity threshold of 0.9, meaning that all smart contracts sharing over 90\% of tokens were discarded.

\subsubsection{Explanation-Guided Fine-Tuning}

Our labeled dataset for Explanation-Guided Fine-Tuning is based on datasets from \cite{liu2023rethinking} and \cite{yu2023pscvfinder}, encompassing various types of SC vulnerabilities. The authors of \cite{liu2023rethinking} manually verified the generated vulnerabilities to ensure their correctness. \cite{yu2023pscvfinder} is the renowned SmartBugs dataset \cite{ferreira2020smartbugs}, on which \cite{wu2021peculiar} annotated reentrancy vulnerabilities. 

We extracted 1,634 smart contracts containing \texttt{call.value} from \cite{yu2023pscvfinder}, and further isolated 379 contracts containing \texttt{delegatecall}, which we manually annotated. To augment our dataset, we crawled verified contracts from Etherscan. The final SFT dataset comprises 3,382 instances of reentrancy vulnerabilities, 1,165 timestamp dependency vulnerabilities, 1,005 integer overflow/underflow vulnerabilities, and 697 delegatecall vulnerabilities.

\subsection{Smart Contract Vulnerability Explanation Annotation} 

We utilized the LLMs Qwen2-72B-Instruct and Mistral-Large-Instruct-2407-123B to generate detailed explanations of smart contract vulnerabilities. These LLMs were chosen primarily because they are open-source, powerful language models that excel in natural language processing and code comprehension, comparable to GPT-4 but at a lower cost. 


To ensure the relevance and accuracy of the generated content, we designed specialized prompts for each vulnerability type, implementing label-guided analysis. These prompts cover common vulnerability types such as reentrancy, timestamp dependence, delegatecall, and integer overflow/underflow. Our prompt design not only guides the models to explain the vulnerabilities but also instructs them to pinpoint the exact locations of these vulnerabilities within the smart contract code, providing specific line numbers and code blocks where the issues occur. For reentrancy vulnerabilities, the prompts guide the models to focus on the use of call.value(), operation order, external calls, access control, and internal function implementation. When analyzing timestamp dependence vulnerabilities, the models are directed to examine the use of block.timestamp or now, time constraints in critical operations, potential for miner manipulation, and the precision of time measurements and their impact on contract logic. For delegatecall vulnerabilities, the prompts require the models to evaluate the use of delegatecall(), context preservation, state variable manipulation, access control, and internal function implementation. Lastly, for integer overflow/underflow vulnerabilities, the models are instructed to check arithmetic operations (especially on uint variables), the use of SafeMath library or Solidity 0.8.x built-in overflow/underflow checks, the use of the 'unchecked' keyword (Solidity 0.8.x or higher), arithmetic operations in critical operations, and type conversion and large number handling. Through this comprehensive and detailed prompt design, we ensure that the generated explanations cover the core elements of each vulnerability type.



To further improve the quality of explanations, we introduce Llama-3.1-70B-Instruct as an evaluation model. This LLM assesses the explanations generated by Qwen2 and Mistral, with evaluation criteria including correctness, completeness, and conciseness, each scored on a scale of 1 to 10. During the scoring process, Llama-3.1-70B-Instruct provides detailed scoring rationales and improvement suggestions. We then select the highest-scoring explanations based on this evaluation for further human review.

In the human expert review stage, we selected 8 PhD students specializing in smart contract vulnerability detection. These experts were divided into 4 groups of 2, with each group responsible for reviewing one specific type of vulnerability. The selected high-scoring explanations from the LLM evaluation undergo careful review by these experts to verify accuracy. If errors, omissions, or unclear points are found, the experts make necessary modifications and additions to further improve the quality of the explanations.

\subsection{Two Stage Post-training for Smart Contract Vulnerability Detection Task}

We propose an innovative two-stage post-training strategy aimed at enhancing the model's understanding of smart contracts, improving vulnerability detection accuracy, and increasing the quality of explanations. In the Smart Contract-Specific Continual Pre-training stage, the model learns to understand the basic structure, syntax, and semantics of smart contracts by minimizing a context-based language model loss function:

\begin{equation}
L_{adapt} = -\sum \log P(x_i | context)
\end{equation}

where $x_i$ represents each token in the input sequence, $context$ represents the surrounding context tokens of $x_i$, and $P(x_i | context)$ is the conditional probability of the model predicting $x_i$. In the case of smart contract analysis, this formula takes on specific meanings: $x_i$ represents each Solidity token in the smart contract code (e.g., 'function', 'uint256', 'require'). The $context$ typically includes function definitions, state variable declarations, and control structures specific to Solidity. $P(x_i | context)$ then becomes the probability of the model correctly predicting Solidity-specific tokens given their surrounding code context. By minimizing this loss, the model learns: (a) Solidity-specific syntax, including contract declarations and function modifiers (e.g., 'payable', 'view'); (b) common smart contract patterns, such as the 'Checks-Effects-Interactions' pattern for preventing reentrancy attacks; (c) EVM-specific operations, including gas considerations and storage vs. memory usage; (d) security-critical Solidity functions, such as 'transfer', 'send', and 'call.value()'. (e) contextual relevance, such as identifying potential security flaws in contract logic, analyzing inter-contract dependencies and interactions, assessing the impact of external calls, and evaluating the correct implementation of access control mechanisms; To maintain the model's general capabilities and prevent catastrophic forgetting, we integrated a substantial amount of general data into the training dataset, including content from mathematics, coding, and linguistics. This approach not only helps the model maintain its understanding of a wide range of knowledge domains but also enhances its generalization ability and robustness in smart contract tasks, reduces the risk of overfitting, and improves the model's performance when dealing with novel or rare contract patterns.

In the Explanation-Guided fine-tuning stage, the model focuses on learning to identify specific types of vulnerabilities and generate corresponding explanations. We adopted a balanced loss function:

\begin{equation}
L_{SFT} = - \sum[\log P(y_g|x;\theta) + \log P(y_d|x;\theta)]/2
\end{equation}

where $x$ represents the input smart contract code, $y_g$ and $y_d$ represent target outputs for the generation task (explanations) and detection task (labels), respectively, and $\theta$ represents all trainable parameters of the model, including attention weights, feedforward layer weights, and embeddings. This loss function ensures that the model is adequately trained on both vulnerability detection and explanation generation tasks. The domain-specific knowledge acquired during the Smart Contract-Specific Continual Pre-training stage lays the foundation for the fine-tuning stage, enabling the model to understand the context and nuances of smart contracts while learning to detect vulnerabilities. 

Through this two-stage strategy, our model first gains a deep understanding of the specific language and structure of the smart contract domain while maintaining diversity in general knowledge. It then focuses on improving the accuracy of vulnerability detection and the quality of explanations.

\subsection{Evaluation of Explanations}

To comprehensively assess the quality of smart contract vulnerability explanations generated by Smart-LLaMA, we follow \cite{wang2023generating} and designed an evaluation framework based on three key dimensions: correctness, completeness, and conciseness. Each dimension is scored on a 4-point Likert scale \cite{joshi2015likert}.

\subsubsection{Evaluation Criteria}
\label{subsubsec:evaluation_criteria}
Our evaluation is based on three key dimensions: Correctness, Completeness, and Conciseness,

\textbf{Correctness}. This criterion evaluates the accuracy of the explanation in terms of reasoning logic and vulnerability localization.

1 - Disagree: Major errors in logic and localization.
2 - Somewhat disagree: Some errors, misses major vulnerabilities.
3 - Somewhat agree: Minor omissions, locates major vulnerabilities.
4 - Agree: Correct logic, accurate identification and localization.

\textbf{Completeness}. This criterion assesses whether the explanation comprehensively covers all potential vulnerability points.

1 - Disagree: Omits multiple key vulnerabilities, superficial explanations.
2 - Somewhat disagree: Identifies some, misses major issues, lacks depth.
3 - Somewhat agree: Covers major vulnerabilities, may miss minor ones.
4 - Agree: Comprehensive identification, detailed explanations for all.

\textbf{Conciseness}. This criterion evaluates whether the explanation is concise and easy to understand and apply quickly.

1 - Disagree: Verbose, key points obscured, difficult to apply.
2 - Somewhat disagree: Somewhat verbose, key info present but unclear.
3 - Somewhat agree: Generally concise, some parts slightly verbose.
4 - Agree: Precise, clear, directly applicable, no redundancy.

\subsubsection{LLM-based Automated Evaluation}

We utilized the advanced LLM Llama-3.1-70B-Instruct for automated evaluation. We carefully designed a set of prompts containing detailed scoring criteria, example scores, and scoring rationales to guide the LLM in evaluating smart contract vulnerability explanations. The LLM scored each explanation generated by Smart-LLaMA on correctness, completeness, and conciseness using a 4-point Likert scale \cite{joshi2015likert}. In addition to providing specific scores, the LLM also offered detailed rationales for each dimension's score. We then developed automated scripts to calculate the distribution of scores for each vulnerability category in each dimension as the final evaluation result.

\subsubsection{Human Evaluation}

To further validate the evaluation results, we invited four experienced smart contract security experts for additional expert evaluation. Each expert spent 8 hours evaluating explanations for one evaluation category, totaling 32 hours of in-depth analysis. The experts used the same 4-point Likert scale \cite{joshi2015likert} as the LLM, scoring each explanation on correctness, completeness, and conciseness. They also provided detailed scoring rationales, improvement suggestions, and overall quality assessments. To ensure consistency, we arranged for 20\% overlapping evaluation samples, allowing different experts to score the same explanation. If the evaluation difference exceeded 1 point on the Likert scale, we discussed and re-evaluated together to reach a consensus. Finally, similar to the LLM-based evaluation, we calculated the distribution of scores for each category in each dimension as the final evaluation result.


\section{Experiments}
\subsection{Research Questions}
To evaluate our proposed Smart-LLaMA approach, we conduct experiments to answer the following research questions:





\begin{itemize}
\item \textbf{RQ1}: How does Smart-LLaMA perform in detecting smart contract vulnerabilities compared to state-of-the-art methods?

\item \textbf{RQ2}: What is the impact of different components in Smart-LLaMA on its overall performance?

\item \textbf{RQ3}: How effective are the explanations generated by Smart-LLaMA in terms of correctness, completeness, and conciseness?

\item \textbf{RQ4}: What insights can be gained from detailed analysis of specific cases in smart contract vulnerability detection using Smart-LLaMA?
\end{itemize}

\subsection{Dataset}



\textbf{Smart Contract-Specific Continual Pre-training:} we employ a dataset derived from the work of Storhaug et al. \cite{storhaug2023efficient}. This dataset comprises 186,397 unique smart contract instances from the Ethereum blockchain, totalling 501.62M tokens. We also augment this dataset with an additional 100,000 instances from various domains, including general code, mathematics, English, and Chinese text, totalling 118.94M tokens. This results in a comprehensive dataset of 286,397 instances, totaling 620.56M tokens.

\textbf{Explanation-Guided Fine-Tuning:} Our Explanation-Guided Fine-Tuning dataset is curated from multiple sources, primarily drawing from Liu et al. \cite{liu2023rethinking} and Yu et al. \cite{yu2023pscvfinder}. We incorporate manually verified vulnerabilities from \cite{liu2023rethinking} and the SmartBugs dataset \cite{ferreira2020smartbugs} with reentrancy vulnerabilities annotated by Wu et al. \cite{wu2021peculiar}. To enrich the dataset, we extracted and manually annotated additional contracts from \cite{yu2023pscvfinder} and Etherscan. The final dataset comprises 3,382 reentrancy, 1,165 timestamp dependency, 1,005 integer overflow/underflow, and 697 delegatecall vulnerability instances, totalling 7.87M tokens.

\textbf{Evaluation:} We utilize the evaluation dataset from \cite{qian2023cross}. This challenging dataset encompasses four major vulnerability types: reentrancy, timestamp dependency, integer overflow/underflow, and delegatecall. The complexity of this dataset is primarily reflected in three aspects: First, for reentrancy vulnerabilities, while most samples contain \texttt{call.value} function calls, not all contracts with this function are vulnerable, requiring models to understand complex access control and execution sequences; Second, identifying timestamp dependency and delegatecall vulnerabilities demands deep contextual understanding, as most contracts contain \texttt{blockstamp} and \texttt{delegatecall}, but the mere presence of these keywords is insufficient to determine the existence of vulnerabilities; Lastly, the integer overflow type includes both overflow and underflow scenarios, increasing the diversity of vulnerability patterns. We removed the unreasonable labels in the original evaluation dataset \cite{qian2023cross}.
\subsection{Baselines}
Our evaluation includes a range of baselines for Smart Contract Vulnerability Detection, representing state-of-the-art approaches in four categories: rule-based, neural network-based, pre-trained model-based, and LLM-based techniques.

Baseline methods, categorized as rule-based techniques, employ predefined heuristics to detect vulnerabilities in smart contracts. This category includes tools such as Mythril \cite{mueller2017mythril}, Osiris \cite{torres2018osiris}, Oyente \cite{luu2016making}, Slither \cite{feist2019slither}, and Smartcheck \cite{tikhomirov2018smartcheck}.

In contrast, neural network-based techniques harness the power of deep learning algorithms to detect vulnerabilities in smart contracts. We consider several cutting-edge methods in this category: GCN \cite{kipf2016semi}, TMP \cite{zhuang2020smart}, AME \cite{liu2021smart}, SMS \cite{qian2023cross} and DMT \cite{qian2023cross}. It should be noted that we faced challenges in reproducing some baselines, and in other cases, our reproduced results differed significantly from the originally reported values. For fairness, we used the higher of our reproduced results and the originally reported figures for each baseline in our comparisons.

Pre-trained models-based techniques, rely on pre-trained models like CodeT5 \cite{wang2021codet5}, CodeBERT \cite{feng2020codebert}, GraphCodeBERT \cite{guo2020graphcodebert} and fine-tuning techniques to identify smart contract vulnerabilities, including Peculiar \cite{wu2021peculiar} and PSCVFinder \cite{yu2023pscvfinder}.

LLM-based techniques, rely on large language models to identify smart contract vulnerabilities, including LLaMA-3.1-8B-Instruct \cite{dubey2024llama}, LLaMA-3.1-70B-Instruct \cite{dubey2024llama}, Qwen2-7B-Instruct \cite{yang2024qwen2}, and Qwen2-72B-Instruct \cite{yang2024qwen2}.



\subsection{Metrics}
To evaluate our proposed model and baseline approaches in vulnerability identification, we employ four widely accepted metrics: Precision, Recall, F1-score, and Accuracy. Precision measures the ratio of correctly identified vulnerabilities to all predicted positives, while Recall calculates the proportion of detected vulnerabilities among all actual vulnerabilities. The F1-score provides a balanced measure by computing the harmonic mean of Precision and Recall. Accuracy assesses overall correctness by calculating the ratio of correct predictions to total cases. These metrics collectively offer a comprehensive assessment of our Smart-LLaMA.

For assessing the quality of vulnerability explanations generated by Smart-LLaMA and baseline models, we utilize three key metrics as discussed in Section~\ref{subsubsec:evaluation_criteria}: Correctness, Completeness, and Conciseness.

\begin{table*}[!ht]
\setlength{\tabcolsep}{5pt} 
    \caption{The performance of our method compared with 16 baselines in terms of Accuracy, Precision, Recall and F1-score. Note: LLM-based techniques use Instruct versions of LLaMA-3.1 and Qwen2 models.}
    \label{TAB1}
    \centering
    
    \begin{tabular}{@{}c@{\hspace{0pt}}|cccc@{\hspace{0pt}}|cccc@{\hspace{0pt}}|cccc@{\hspace{0pt}}|cccc@{}}
        \toprule
        \raisebox{-1\height}{\centering Methods} & \multicolumn{4}{c|}{Reentrancy} & \multicolumn{4}{c|}{Timestamp Dependency} & \multicolumn{4}{c|}{Overflow/Underflow} & \multicolumn{4}{c}{Delegatecall}\\ 
        & A(\%) & P(\%) & R(\%) & F1(\%) & A(\%) & P(\%) & R(\%) & F1(\%) & A(\%) & P(\%) & R(\%) & F1(\%) & A(\%) & P(\%) & R(\%) & F1(\%)\\ 
        \midrule
        Mythril & 54.21 & 33.75 & 73.97 & 46.35 & 43.27 & 44.63 & 44.13& 44.38 & 37.45 & 25.30 & 46.67 & 32.81 & 60.71 & 42.99 & 74.19 & 54.44\\ 
        Osiris & 30.77 & 26.80 & 91.78 & 41.49 & 51.03 & 52.50 & 36.63 & 43.15 & 61.90 & 45.33 & 75.56 & 56.67 & -- & -- & -- & --\\ 
        Oyente & 68.50 & 42.35 & 49.32 & 45.57 & 53.98 & 66.67 & 18.60 & 29.09 & 72.53 & 60.87 & 46.67 & 52.83 & 64.71 & 40.00 & 31.58 & 35.29\\ 
        Slither & 37.73 & 16.08 & 31.51 & 21.30 &57.02 & 56.50 & 70.39 & 62.69 & 50.91 & 32.28 & 45.56 & 37.79 & 52.04 & 39.04 & 91.94 & 54.81\\ 
        Smartcheck & 43.22 & 30.10 & 84.93 & 44.44 & 51.00 & 57.14 & 17.88 & 27.23 & 52.00 & 31.25 & 38.89  & 34.65 & 54.08 & 32.93 & 43.55 & 37.50\\ 
        \midrule
        GCN & 73.21 & 74.47 & 73.18 & 73.82 & 75.91 & 74.93 & 77.55 & 76.22 & 67.53 & 69.52 & 70.93 & 70.22 & 65.76 & 69.01 & 69.74 & 69.37\\ 
        TMP & 76.45 & 76.04 & 75.30 & 75.67 & 78.84 & 78.68 & 76.09 & 77.36 & 70.85 & 70.26 & 69.47 & 69.86 & 69.11 & 68.18 & 70.37 & 69.26\\ 
        AME & 81.06 & 79.62 & 78.45 & 79.03 & 82.25 & 81.42 & 80.26 & 80.84 & 73.24 & 71.36 & 71.59 & 71.47 & 72.85 & 70.25 & 69.40 & 69.82\\
        SMS & 83.85 & 79.46 & 77.48 & 78.46 & 89.77 & 89.15 & 91.09  & 90.11 & 79.36 & 78.14 & 72.98 & 75.47 & 78.82 & 76.97 & 73.69 & 75.29\\ 
        DMT & 89.42 & 83.62 & 81.06 & 82.32 & 94.58 & 93.60 & 96.39 & 94.97 & 85.64 & 85.44 & 74.32 & 79.49 & 82.76 & 84.61 & 77.93 & 81.13\\ 
        \midrule
        Peculiar & 58.72 & 35.23 & 48.44 & 40.84 & 69.08 & 77.31 & 71.12 & 74.18 & 75.91 & 64.29 & 60.00 & 62.07 & 88.46 & 77.36 & 82.00 & 79.61\\ 
        PSCVFinder & 58.26 & 35.16 & 50.00 & 41.32 & 40.10 & 52.00 & 50.39 & 51.18 & 51.09 & 33.58 & 50.00 & 40.18 & 89.56 & 79.25 & 84.00 & 81.57\\ 
        \midrule
        LLaMA-3.1-8B & 36.24 & 29.73 & 85.94 & 44.18 & 57.00 & 67.24 & 60.47 & 63.67 & 55.11 & 38.93 & 64.44 & 48.54 & 63.74 & 38.89 & 56.00 & 45.90\\ 
        Qwen2-7B & 30.28 & 29.63 & 100.00 & 45.71 & 67.63 & 71.83 & 79.07 & 75.28 & 66.42 & 47.62 & 22.22 & 30.30 & 65.93 & 44.23 & 92.00 & 59.74\\ 
        LLaMA-3.1-70B & 29.36 & 29.36 & 100.00 &45.39 & 62.32 & 62.32 & 100.00 & 76.79 & 77.74 & 64.08 & 73.33 & 68.39 & 65.93 & 44.55 & 98.00 & 61.25\\ 
        Qwen2-72B & 35.78 & 31.37 & 100.00 & 47.76 & 69.08 & 67.38 & 97.67 & 79.75 & 75.55 & 64.94 & 55.56 & 59.88 & 74.73 & 52.08 & 100.00 & 68.49\\
        \textbf{Smart-LLaMA} & \textbf{93.12} & \textbf{87.69} & \textbf{89.06} & \textbf{88.37} & \textbf{95.17} & \textbf{95.42} & \textbf{96.90} & \textbf{96.15} & \textbf{89.78} & \textbf{79.25} & \textbf{93.33} & \textbf{85.71} & \textbf{94.51} & \textbf{95.45} & \textbf{84.00} & \textbf{89.36}\\
        \bottomrule
    \end{tabular}
\end{table*}

\subsection{Implementation Details}

We perform Smart Contract-Specific Continual Pre-training and Explanation-Guided Fine-Tuning using LlamaFactory~\cite{zheng2024llamafactory} and DeepSpeed~\cite{rasley2020deepspeed} with \texttt{fp16} enabled.
We calculate loss with cross-entropy and optimize parameters using AdamW~\cite{adamw} with $\beta$=$(0.9, 0.99)$ and $\epsilon$=$1$e-$8$. For all our models, we employ full parameter fine-tuning and continual pre-training. During Smart Contract-Specific Continual Pre-training, we set the batch size to $64$ per device, gradient accumulation steps to $16$, epochs to $2$, learning rate to $1$e-$5$ with cosine decay, warmup steps to $0$, cutoff length to $2048$, and save steps to $500$. During Explanation-Guided Fine-Tuning, we set the batch size to $8$ per device, gradient accumulation steps to $8$, epochs to $3$, learning rate to $1$e-$5$ with cosine decay, warmup steps to $0$, cutoff length to $2048$, and save steps to $50$. All models were trained on a server equipped with 8 NVIDIA GeForce RTX H800 GPUs, each with 80GB memory. For evaluating our Smart-LLaMA, we use greedy decoding with do\_sample set to \texttt{false}. Peculiar \cite{wu2021peculiar} originally only detects reentrancy vulnerabilities, while PSCVFinder \cite{yu2023pscvfinder} detects reentrancy and timestamp dependency vulnerabilities. We extended both tools to detect two additional vulnerability types using the same code.
\subsection{Experimental Results}
In this section we present experimental results to answer the research question.

1) RQ1: To answer this question, we compared Smart-LLaMA's performance against 16 baseline methods across four different vulnerability types: Reentrancy, Timestamp Dependency, Overflow/Underflow, and Delegatecall. The results are presented in Table \uppercase\expandafter{\romannumeral2}.

Smart-LLaMA demonstrated superior performance across all vulnerability types, consistently outperforming state-of-the-art (SOTA) methods. For Reentrancy vulnerabilities, Smart-LLaMA achieved the highest scores with an F1-score of 88.37\% and accuracy of 93.12\%, surpassing the previous best performer DMT by 7.35\% in F1-score and 4.14\% in accuracy. In detecting Timestamp Dependency vulnerabilities, Smart-LLaMA again led with the highest F1-score of 96.15\% and accuracy of 95.17\%, outperforming DMT by 1.24\% in F1-score and 0.62\% in accuracy. For Overflow/Underflow vulnerabilities, Smart-LLaMA maintained its lead with the highest F1-score of 85.71\% and accuracy of 89.78\%, surpassing DMT by 7.82\% in F1-score and 4.83\% in accuracy. Finally, for Delegatecall vulnerabilities, Smart-LLaMA showed strong performance with the highest F1-score of 89.36\% and accuracy of 94.51\%, exceeding the previous best performer PSCVFinder by 9.55\% in F1-score and 5.53\% in accuracy. Notably, Smart-LLaMA consistently outperformed other popular methods such as Mythril, Oyente, Securify, and Slither across all vulnerability types, and showed significant improvements over more recent LLM-based approaches like LLaMA-3.1-8B and Qwen2-7B in most metrics.


The superior performance of Smart-LLaMA can be attributed to three key factors. Our comprehensive dataset provided rich learning resources, covering various vulnerability types and complex scenarios. The Smart Contract-Specific Continual Pre-training and Explanation-Guided Fine-Tuning enabled the model to deeply understand smart contract structures and the detection process, significantly enhancing its detection capabilities.

\begin{tcolorbox}
\textbf{[RQ1]}: Smart-LLaMA consistently outperformed 16 state-of-the-art baseline methods across all four types of vulnerabilities (reentrancy, timestamp dependency, integer overflow/underflow, and delegatecall).
\end{tcolorbox}


2) RQ2: Our ablation study elucidated the crucial roles of Explanation-Guided Fine-Tuning (EGFT) and Smart Contract-Specific Continual Pre-Training (SCPT) in the Smart-LLaMA architecture. The base model represents Smart-LLaMA's full performance. "w/o EGFT" indicates the model trained only with labeled data for Supervised Fine-Tuning, without the vulnerability explanations. "w/o SCPT" represents the model without Smart Contract-Specific Continual Pre-Training, and "w/o Both" shows the performance without both components.

\begin{table}[htbp]
\centering
\caption{Ablation Study of Smart-LLaMA.}
\begin{tabular}{|c|c|c|c|c|c|}
\hline
Types & Metric & Base & w/o EGFT & w/o SCPT & w/o Both \\
\hline
\multirow{2}{*}{RE} & Acc(\%) & 93.12 & 71.10  & 84.86  & 71.10 \\
\cline{2-6}
 & F1(\%) & 88.37 & 5.97 & 68.57  & 5.97 \\
\hline
\multirow{2}{*}{TD} & Acc(\%) & 95.17 & 54.59 & 83.09 & 39.13 \\
\cline{2-6}
 & F1(\%) & 96.15 & 48.91 & 87.46 & 16.00 \\
\hline
\multirow{2}{*}{IO} & Acc(\%) & 89.78 & 59.85  & 79.93 & 58.76 \\
\cline{2-6}
 & F1(\%) & 85.71 & 45.00 & 57.36 & 33.92 \\
\hline
\multirow{2}{*}{DE} & Acc(\%) & 94.51 & 67.58  & 86.26 & 62.09 \\
\cline{2-6}
 & F1(\%) & 89.36  & 35.16 & 67.53 & 12.66 \\
\hline
\end{tabular}
\label{tab:vulnerability_detection}
\end{table}

For reentrancy vulnerabilities, the complete Smart-LLaMA achieved 93.12\% accuracy and 88.37\% F1 score. Removing EGFT significantly reduced performance (71.10\% Acc, 5.97\% F1), while eliminating SCPT has a lesser impact (84.86\% Acc, 68.57\% F1). The drastic drop in F1 score without EGFT (from 88.37\% to 5.97\%) indicated that the detailed reasoning process in Explanation-Guided Fine-Tuning is particularly important for the LLM to understand and identify the complex logic of reentrancy vulnerabilities. 

In detecting timestamp dependency vulnerabilities, Smart-LLaMA achieved an impressive 95.17\% accuracy and 96.15\% F1 score. Removing EGFT severely impacts performance (54.59\% Acc, 48.91\% F1), while omitting SCPT has a milder effect (83.09\% Acc, 87.46\% F1). Notably, excluding both components led to a dramatic performance decline (39.13\% Acc, 16.00\% F1), highlighting their synergistic importance.

For integer overflow vulnerabilities, the complete model attained 89.78\% accuracy and 85.71\% F1 score. EGFT's removal caused a substantial drop (59.85\% Acc, 45.00\% F1), while SCPT's absence had a moderate impact (79.93\% Acc, 57.36\% F1). Similarly, for delegate call vulnerabilities, Smart-LLaMA achieved 94.51\% accuracy and 89.36\% F1 score. Again, removing EGFT resulted in a more significant decline (67.58\% Acc, 35.16\% F1) compared to removing SCPT (86.26\% Acc, 67.53\% F1).


We can summarize the following general patterns: (1) The complete model performs best across all vulnerability types, demonstrating the importance of both modules in Smart-LLaMA; (2) The impact of EGFT is generally more significant than that of SCPT, indicating that detailed reasoning processes are particularly important for understanding and identifying complex vulnerability logic; (3) While SCPT has a relatively smaller impact, it still positively contributes to the performance by providing broad domain knowledge to support the model's contextual understanding and reasoning depth when analyzing smart contracts.


\begin{tcolorbox}
\textbf{[RQ2]}: Explanation-Guided Fine-Tuning (EGFT) and Smart Contract-Specific Continual Pre-Training (SCPT) both significantly enhance Smart-LLaMA's performance. EGFT shows a more substantial impact, especially on F1 scores.
\end{tcolorbox}

3) RQ3: Table \uppercase\expandafter{\romannumeral4} presents the quality assessment results of explanations generated by the Smart-LLaMA method for smart contract vulnerability detection tasks. The evaluation comprises two parts: LLM evaluation and human evaluation, each assessing three dimensions: Correctness, Completeness, and Conciseness.

\begin{table}[htbp]
\centering
\caption{Ratings of Correctness, Completeness, and Conciseness}
\label{tab:ratings}
\setlength{\tabcolsep}{3.5pt} 
\begin{tabular}{|l|cccc|cccc|cccc|}
\hline
\multirow{2}{*}{} & \multicolumn{4}{c|}{Correctness} & \multicolumn{4}{c|}{Completeness} & \multicolumn{4}{c|}{Conciseness} \\
\cline{2-13}
& 1 & 2 & 3 & 4 & 1 & 2 & 3 & 4 & 1 & 2 & 3 & 4 \\
\hline
\multicolumn{13}{|c|}{LLM Evaluation} \\
\hline
Baseline & 140 & 110 & 86 & 545 & 21  & 187 & 221 & 452 & 9 & 171 & 474 & 227 \\
Ours & 28 & 40 & 25 & 788 & 4 & 60 & 132 & 685 & 0 & 11 & 293 & 577 \\
\hline
\multicolumn{13}{|c|}{Human Evaluation} \\
\hline
Baseline & 33 & 245 & 310 & 293 & 26 & 209 & 449 & 197 & 11 & 147 & 569 & 154 \\
Ours & 16 & 79 & 174 & 612 & 13 & 95 & 270 & 503 & 1 & 14 & 288 & 578 \\
\hline
\end{tabular}
\end{table}


In both LLM and human evaluations, Smart-LLaMA significantly outperformed the baseline (LLaMA-3.1-8B-Instruct). In the LLM evaluation for correctness, Smart-LLaMA achieved a score of 4 (agree) in 89.4\% of cases and 3 (somewhat agree) in 2.8\%, compared to the baseline's 61.9\% and 9.8\%. For completeness, Smart-LLaMA scored 4 in 77.8\% of cases and 3 in 15.0\%, substantially higher than the baseline's 51.3\% and 25.1\%. In terms of conciseness, Smart-LLaMA excelled with 65.5\% scoring 4 and 33.3\% scoring 3, versus the baseline's 25.8\% and 53.8\%. evaluation results further corroborated Smart-LLaMA's advantages. For correctness, Smart-LLaMA received a score of 4 in 69.5\% of cases and 3 in 19.8\%, notably higher than the baseline's 33.3\% and 35.2\%. In completeness, Smart-LLaMA achieved a score of 4 in 57.1\% of cases and 3 in 30.6\%, compared to the baseline's 22.4\% and 51.0\%. Regarding conciseness, Smart-LLaMA again demonstrated superior performance, with 65.6\% scoring 4 and 32.7\% scoring 3, while the baseline achieved 17.5\% and 64.6\% respectively.

These results clearly indicated that the Smart-LLaMA method generates more accurate, comprehensive, and concise explanations for smart contract vulnerability detection. 

Notably, human evaluation scores were consistently lower than those from LLM evaluation, highlighting a potential gap between automatically generated explanations and human expert expectations. This observation suggested a direction for future research: further optimization of generative models to produce explanations that better align with human.




\begin{tcolorbox}
\textbf{[RQ3]}: Smart-LLaMA has demonstrated its capability to generate explanations for smart contract vulnerability that are correct, complete, and concise, as validated by both LLM evaluation and human evaluation.

\end{tcolorbox}

4) RQ4: We demonstrated the superiority of Smart-LLaMA in smart contract vulnerability detection through a specific case study. As illustrated in Figure 4, we compared Smart-LLaMA with other methods such as Slither, PSCVFinder, and Peculiar.

\begin{figure}[!h]
\centerline{\includegraphics[width=0.5\textwidth]{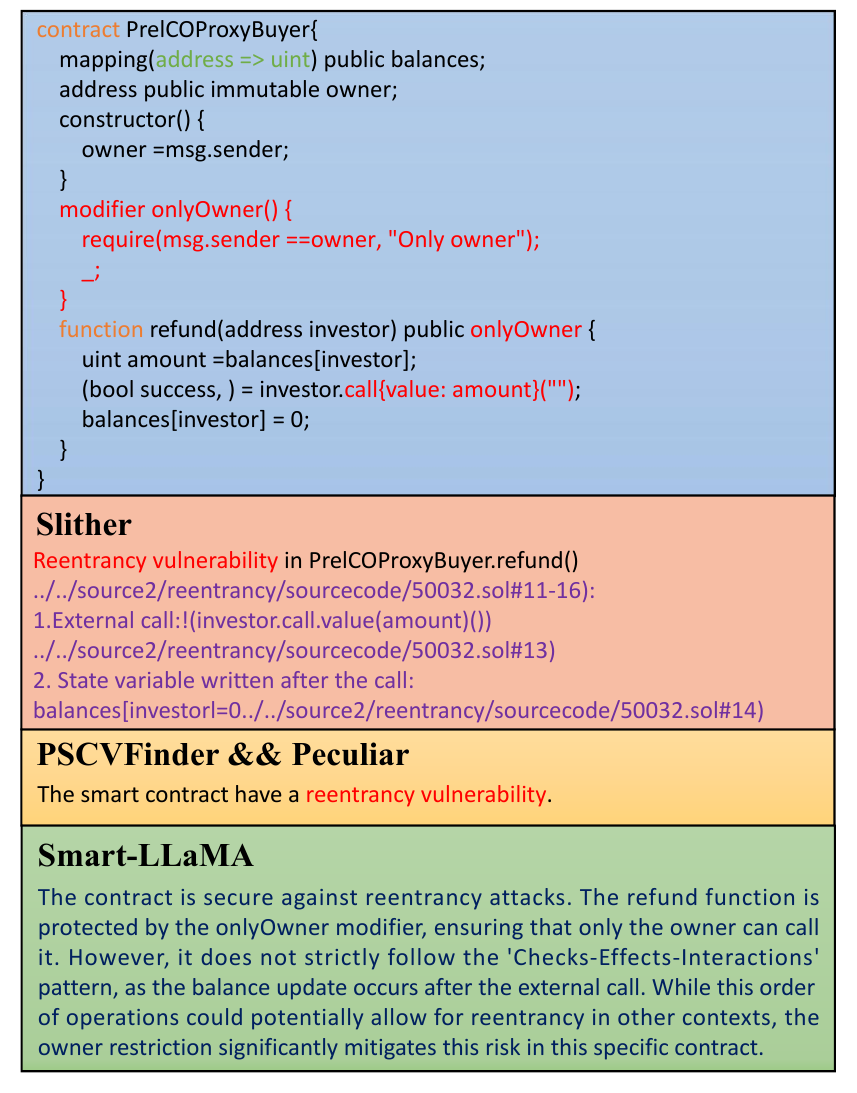}}
\caption{A Case Study of Smart Contract Vulnerability Detection Using Smart-LLaMA.}
\label{fig}
\end{figure}

This case study highlighted the advantages of our two-stage post-training method. Slither, PSCVFinder, and Peculiar all incorrectly identified a reentrancy vulnerability, failing to accurately assess the contract's security. These tools overlooked crucial security features, particularly the role of the onlyOwner modifier. In contrast, Smart-LLaMA successfully provided the only correct assessment. It accurately recognized that the contract is secure against reentrancy attacks and offered an in-depth analysis. Smart-LLaMA's smart contract-specific pre-training enabled it to deeply understand the contract structure, especially in accurately identifying the critical role of the onlyOwner modifier in protecting the refund function.

Smart-LLaMA's explanation-guided fine-tuning further enhanced its analytical capabilities. Although it noted that the balance update occurs after the external call (which typically could lead to reentrancy issues), Smart-LLaMA accurately assessed that in this specific case, the owner restriction (onlyOwner) effectively eliminates the reentrancy risk. This nuanced analysis demonstrates Smart-LLaMA's ability to consider multiple security factors and context.


\begin{tcolorbox}
\textbf{[RQ4]}: Compared to other methods, Smart-LLaMA not only accurately identifies the contract's security features but also provides in-depth contextual analysis, thus demonstrating clear advantages in complex smart contract security evaluations.
\end{tcolorbox}




\section{Related Work}
Smart contract vulnerability detection research has evolved significantly, transitioning from traditional rule-based methods to advanced machine learning and LLM-based approaches. Early studies employed various program analysis techniques: Oyente \cite{luu2016making} pioneered the use of symbolic execution to uncover four types of vulnerabilities by exploring different execution paths. Tools like Mythril \cite{mueller2017mythril} and SmartCheck \cite{tikhomirov2018smartcheck} relied on pattern matching against predefined vulnerability patterns. Securify \cite{tsankov2018securify} utilized formal verification with custom-designed compliance and violation patterns. Others, like Osiris \cite{torres2018osiris} and Manticore \cite{mossberg2019manticore}, combined multiple techniques such as symbolic execution and taint analysis.

The field then progressed to leverage deep learning models. BLSTM-ATT \cite{qian2020towards} introduced a bidirectional LSTM with an attention mechanism for reentrancy vulnerability detection. SmartEmbed \cite{gao2019smartembed} used deep learning to measure similarity to contracts with known vulnerabilities. Graph neural networks (GNNs) gained popularity due to their ability to capture structural information: works by Zhuang et al. \cite{zhuang2020smart} and Liu et al. \cite{liu2021smart} utilized GNNs to represent contract semantics, with Liu et al. combining GNNs with expert patterns. Recent advancements include pre-training models like Peculiar \cite{wu2021peculiar}, which focused on improving generalization, and ReVulDL \cite{zhang2022reentrancy}, which utilized a graph-based pre-training model to capture propagation chain relationships. Innovative approaches like PSCVFinder \cite{yu2023pscvfinder} employed prompt-tuning to bridge the gap between pre-training and downstream tasks, while Qian et al. \cite{qian2023cross} introduced cross-modality mutual learning to harness complementary information from bytecode and source code.

Most recently, the potential of Large Language Models (LLMs) in smart contract security has been explored. Studies by Chen et al. \cite{chen2023chatgpt} and David et al. \cite{david2023you} evaluated LLMs on real-world datasets, revealing both their potential and challenges, particularly high false positive rates. Hu et al. \cite{hu2023large} investigated new perspectives for LLMs in vulnerability detection, exploring their reasoning capabilities. Sun et al. \cite{sun2024gptscan} introduced GPTScan, an innovative approach combining GPT with program analysis. GPTScan breaks down each logic vulnerability type into scenarios and properties, utilizes GPT to match candidate vulnerabilities, and then confirms them through static analysis. These works demonstrate the promising future of LLMs in enhancing smart contract security, while also highlighting areas for improvement.


Despite these advancements, existing approaches still face limitations. Most works rely on limited datasets. LLM-based methods struggle with domain-specific knowledge, leading to high false positive rates. Additionally, many approaches, especially deep learning-based methods, cannot provide detailed and high-quality explanations for detected vulnerabilities.

\section{Threats to Validity}



\textbf{Internal Validity}: After supervised fine-tuning, Smart-LLaMA may generate repetitive outputs. In our Smart-LLaMA model, we implemented truncation to only retain the first part of the output. This phenomenon could be attributed to the model's tendency to reinforce certain patterns during supervised fine-tuning, potentially leading to redundant information. In future work, we plan to enhance our handling of repetitive outputs, possibly through more sophisticated post-processing techniques or improved fine-tuning strategies.

\textbf{External Validity}: Smart-LLaMA requires a substantial amount of labeled training data. Our data generation process involved two LLMs for initial generation, a third model for voting, and human verification. However, this process cannot guarantee that every reasoning step in the verified SFT data is entirely correct. Additionally, due to resource constraints, we were unable to expand our approach to cover a wider range of vulnerability types. In future research, we aim to address these limitations by exploring more efficient data labeling techniques and expanding our coverage to include a broader spectrum of smart contract vulnerabilities.

\section{Conclusion}


In this paper, we propose Smart-LLaMA, a smart contract vulnerability detection approach based on LLMs. Unlike prior work, Smart-LLaMA novelly combines smart contract-specific continual pre-training with explanation-guided fine-tuning to enhance the performance of smart contract vulnerability detection. Furthermore, we introduce a comprehensive dataset with detailed explanations to improve the model's understanding of vulnerabilities. Our approach significantly outperforms state-of-the-art methods across four major vulnerability types. 

\bibliographystyle{IEEEtran}
\balance
\bibliography{IEEEabrv,References}

\end{document}